\documentclass[journal,twoside, a4paper]{IEEEtran}
\usepackage{multicol}
\usepackage{makecell}
\usepackage{color}
\usepackage{amssymb}
\usepackage{balance}
\usepackage{rotating}
\usepackage[T1]{fontenc}
\usepackage{lscape}
\usepackage{dblfloatfix}
\usepackage{xcolor}
\usepackage{subcaption}
\usepackage[ruled,vlined]{algorithm2e}
\usepackage[autostyle]{csquotes}
\usepackage{tikz}
\usepackage{acro}
\usepackage{siunitx}
\usepackage{enumitem}
\usepackage{multirow}
\usepackage{cite}
\usepackage{amsmath,amssymb,amsfonts}
\usepackage{tikz}
\usepackage{float}
\usepackage{algorithmic}
\usepackage{graphicx}
\usepackage{textcomp}
\usepackage{xcolor}
\usepackage{longtable}
\usepackage{url}
\usepackage{booktabs}
\usepackage{comment}
\def\BibTeX{{\rm B\kern-.05em{\sc i\kern-.025em b}\kern-.08em
    T\kern-.1667em\lower.7ex\hbox{E}\kern-.125emX}}

\ifCLASSINFOpdf
\else
 
\fi

\usepackage[a4paper, total={180mm,257mm}]{geometry}

\begin{document}

\title{BreathAI: Transfer Learning-Based Thermal Imaging for Automated Breathing Pattern Recognition}

\author{\IEEEauthorblockN{
Hamza Kheddar\IEEEauthorrefmark{1}, 
Yassine Himeur\IEEEauthorrefmark{2},  
Abbes Amira\IEEEauthorrefmark{3,4} 
}\\
\IEEEauthorblockA{\IEEEauthorrefmark{1} 
LSEA Laboratory, Department of Electrical Engineering, 
University of Medea, Medea, Algeria; Email: kheddar.hamza@univ-medea.dz} \\

\IEEEauthorblockA{\IEEEauthorrefmark{2} 
College of Engineering and Information Technology, 
University of Dubai, Dubai 14143, UAE; Email: yhimeur@ud.ac.ae} \\

\IEEEauthorblockA{\IEEEauthorrefmark{3} 
College of Computing and Informatics, 
University of Sharjah, Sharjah, UAE; Email: aamira@sharjah.ac.ae} \\

\IEEEauthorblockA{\IEEEauthorrefmark{4} 
Institute of Artificial Intelligence, 
De Montfort University, Leicester, United Kingdom}
}

\maketitle
\thispagestyle{empty}
\pagestyle{empty}
\begin{abstract}
This study presents an Adaptive Transfer Learning and Thresholding-based Deep Learning Model (ATL-TDLM) for automated breathing pattern recognition using thermal imaging. Unlike conventional methods that rely on sound-based respiratory data, our approach leverages hierarchical deep feature extraction and adaptive multi-thresholding (AMT) to enhance feature segmentation. The model integrates knowledge distillation-based fine-tuning (KD-FT) to optimize learning transfer and contrastive representation learning (CRL) to improve inter-class separability between inhalation (INH) and exhalation (EXH) phases. The ATL-TDLM framework achieves an accuracy of 98.8\%, significantly outperforming state-of-the-art models while ensuring computational efficiency. This approach has potential applications in respiratory disorder detection, including sleep apnea and asthma monitoring.
\end{abstract}
\begin{IEEEkeywords}
Breathing disorders, Thermal imaging, Transfer learning, Knowledge distillation, Breathing classification, Fine-tuning
\end{IEEEkeywords}
\section{Introduction}
\label{sec:intro}


Breathing disorders encompass a range of conditions affecting the respiratory system, hindering normal breathing processes. Conditions like sleep apnea disrupt regular breathing patterns during sleep, leading to intermittent pauses \cite{salih2022deep}. Asthma involves recurrent airway inflammation, causing wheezing and shortness of breath. Chronic obstructive pulmonary disease (COPD) includes conditions like chronic bronchitis and emphysema, resulting in persistent airflow limitation. Additionally, pneumonia, characterized by lung inflammation due to infection, impairs efficient breathing. These disorders vary in severity and symptoms but commonly share the disruption of normal respiratory functions, underscoring the importance of timely diagnosis and management for optimal respiratory health \cite{shakhih2019assessment}.

Subjective experience significantly impacts medical image analysis, often introducing variability in respiratory disorder assessments. While computed tomography (CT) imaging serves as a valuable diagnostic tool \cite{habchi2023ai}, its application in breathing disorder detection remains limited due to cost, accessibility, and radiation exposure concerns. Additionally, existing deep learning (DL)-based approaches for respiratory monitoring predominantly rely on sound recordings or sensor-based techniques \cite{qin2021deep, topaloglu2023explainable, 2023towards}, which may be susceptible to ambient noise interference and require specialized equipment.

To address these limitations, we propose an Adaptive Transfer Learning and Thresholding-based Deep Learning Model (ATL-TDLM), designed to enhance breathing pattern recognition through thermal imaging analysis. Our method introduces key innovations that overcome challenges in respiratory phase differentiation, particularly between mid-inhalation (mid-INH) and mid-exhalation (mid-EXH) states. The main contributions of this study are as follows:

\begin{itemize}
    \item Thermal Imaging for Contactless Breathing Detection: Unlike sound-based or sensor-dependent methods, our approach leverages infrared thermal images (TIs) to capture temperature variations in inhalation (INH) and exhalation (EXH), enabling real-time, non-contact monitoring.

    \item Advanced Feature Extraction and Adaptation: We integrate Adaptive Multi-Thresholding (AMT) for dynamic edge detection and data augmentation, along with multi-stage feature extraction (MSFE) using ResNet50 to enhance representation while reducing overfitting and computation.

    \item Robust Classification and Learning Strategies: We employ Knowledge Distillation Fine-Tuning (KD-FT) for domain adaptation, Contrastive Representation Learning (CRL) to refine feature separability, and a hybrid classification approach (SVM and RF) for improved accuracy and interpretability.
\end{itemize}

The remainder of the paper is structured as follows: Section \ref{sec:rw} provides a summary of the most relevant related work, as well as existing breathing assessment techniques. Section \ref{sec:pdm} comprehensively explores the proposed approach. Experiments and research findings will be detailed in Section \ref{sec:exp}. The paper concludes with a summary in Section \ref{sec:conc}.

\section{Related work}
\label{sec:rw}

Breathing disorders significantly impact human health, prompting extensive research on their prediction and detection. Various studies utilize deep learning (DL) and machine learning (ML) techniques for improved diagnosis. Salih et al. \cite{salih2022deep} developed a CNN-LSTM model for asthma and bronchitis classification in preschoolers, enhancing diagnostic efficiency. Qin et al. \cite{qin2021deep} evaluated small airway function in asthmatic children using CT data and DL, showing oral corticosteroids’ efficacy over inhalation therapy via a ResNet model. Yahyaoui et al. \cite{yahyaoui2021deep} explored ML models for asthma and pneumonia detection, achieving 95\% accuracy with KNN and 94.3\% with DNN. Bhat et al. \cite{bhat2021machine} introduced a smartphone-based asthma risk prediction system integrating multiple sensors. Alsaad et al. \cite{alsaad2019interpreting} highlighted the interpretability of DL models in asthma prediction using electronic health records. Topaloglu et al. \cite{topaloglu2023explainable} proposed an attention network with ML for precise asthma detection, utilizing Grad-CAM for explainability.

For sleep apnea detection, Li et al. \cite{li2023deep} employed a ResNet model on EEG and airflow signals, achieving 87.7\% accuracy. Wang et al. \cite{wang2023sleep} compared four neural network models, finding 1D-CNN-LSTM and 2D-CNN-LSTM superior, with accuracy exceeding 83\%. Breathing assessment methods vary. Respiratory belts and spirometry are common, but studies on precise inhalation (INH) and exhalation (EXH) timing remain limited \cite{doheny2023estimation}. Techniques like bioacoustics, strain gauge belts, and respiratory inductive plethysmography (RIP) offer accurate measurements but have practical limitations \cite{kristiansen2023clinical}. ECG-based respiration monitoring is susceptible to movement artifacts \cite{salari2022detection}. Shakhih et al. \cite{shakhih2019assessment} proposed infrared thermal imaging (ITI) for tracking three breathing patterns, but distinguishing sub-states like mid-INH and mid-EXH remains a challenge.

\begin{figure}
    \centering
    \includegraphics[width=0.5\textwidth]{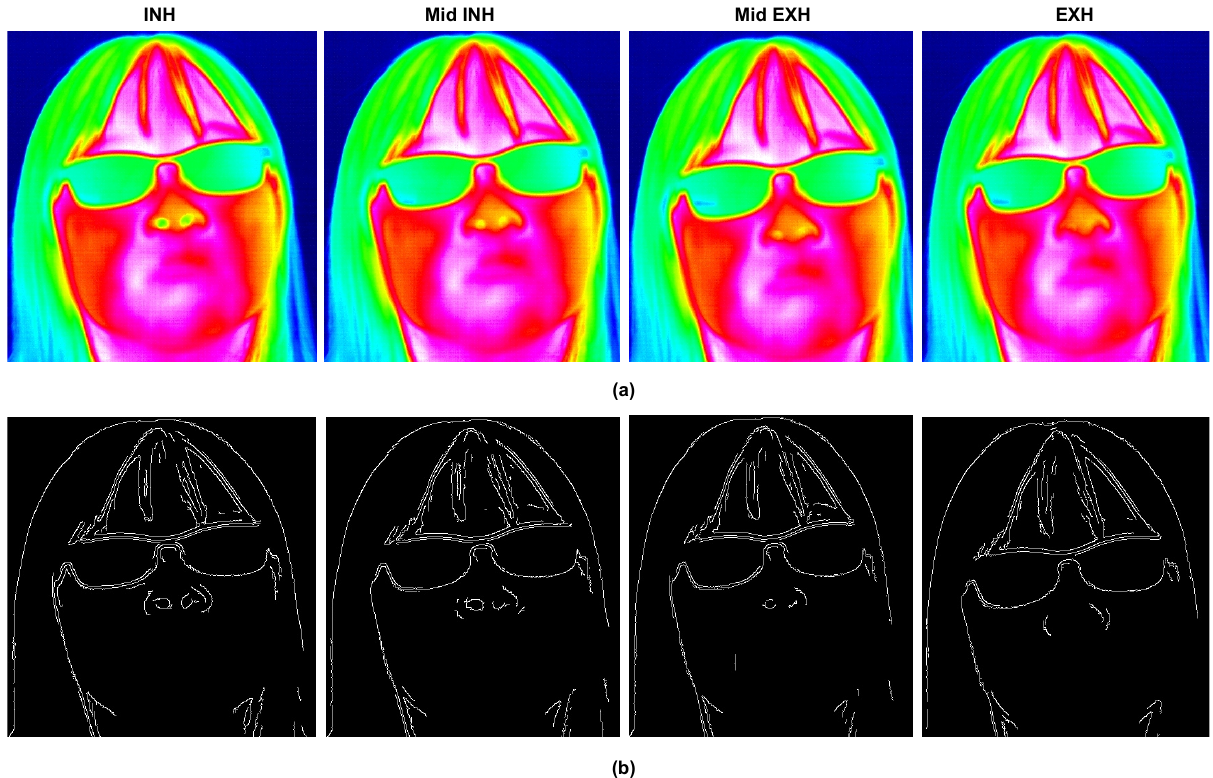}
    \caption{Example of used ITIs encompassing status INH, mid INH, mid EXH, and EXH: (a) Original ITIs (b) Thresholded ITIs. }
    \label{fig:thr}
\end{figure}

\section{Proposed Detection Method}
\label{sec:pdm}

\begin{figure}
    \centering
    \includegraphics[scale=0.7]{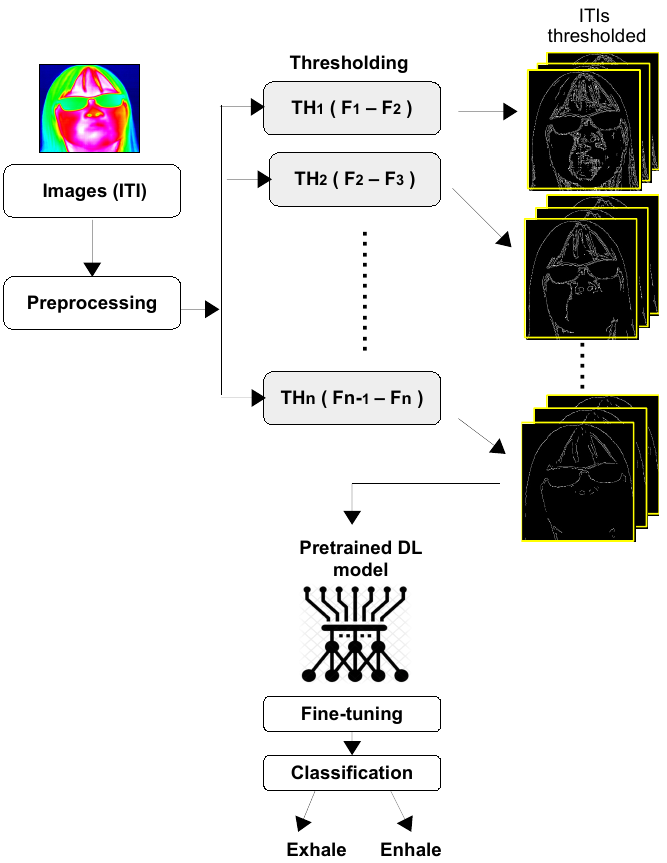}
    \caption{Illustration of the proposed ATL-TDLM framework.}
    \label{fig:scheme}
\end{figure}

The ATL-TDLM framework efficiently classifies breathing patterns, focusing on the challenging mid-INH and mid-EXH states. It aims to enhance classification accuracy by reducing misclassifications, avoiding random guessing, and mitigating model bias. The model integrates Multi-Stage Feature Extraction, utilizing hierarchical deep features from pre-trained convolutional layers. Adaptive Multi-Thresholding (AMT) dynamically refines edge detection, improving inhalation and exhalation clarity. Knowledge Distillation-Based Fine-Tuning (KD-FT) minimizes inter-class ambiguity by optimizing knowledge transfer. Contrastive Representation Learning (CRL) enhances feature separability, preventing confusion between mid-states. Together, these techniques improve generalization and classification accuracy, making ATL-TDLM a robust, scalable, and efficient thermal imaging-based breathing pattern recognition solution. The proposed method follows a structured pipeline, illustrated in Figure~\ref{fig:scheme}, and consists of the following key stages.

\subsection{Preprocessing and Data Augmentation}
To optimize feature representation, the input images undergo multi-level preprocessing:
\begin{itemize}
    \item Gray-Scale Transformation: Images \( X_i \) are converted to single-channel intensity matrices \( G(X_i) \) with values \( p_{x,y} \in [0, 255] \).
    \item Adaptive Multi-Thresholding (AMT): An adaptive thresholding function \( T(F_{\text{low}}, F_{\text{high}}) \) is applied for robust edge detection:
\begin{equation}
\text{Pixel}_{(x,y)} =
\begin{cases} 
  1, & \text{if } G(X_i) \geq F_{\text{high}} \\
  0, & \text{if } G(X_i) < F_{\text{low}} \\
  \text{Edge}, & \text{otherwise}
\end{cases}   
\end{equation}
\item Data Augmentation: Each preprocessed image is transformed into \( K \) augmented versions using rotation, scaling, and brightness adjustments:
\begin{equation}
X_i^{(k)} = \mathcal{T}_k(X_i), \quad k \in \{1,2, \dots, K\}
\end{equation}
where \( \mathcal{T}_k \) represents a transformation function applied to the original image.
\end{itemize}

\subsection{Hierarchical Feature Extraction with Adaptive Transfer Learning}
\label{ssec:subhead}
Deep hierarchical feature extraction is performed using a Multi-Level Convolutional Feature Extractor (MCFE), which combines feature maps from different layers of a pre-trained model \( \mathbb{M}_S \) (ResNet50) with a target model \( \mathbb{M}_T \).

\subsubsection{Knowledge Distillation-Based Fine-Tuning}
The feature transfer function is defined as:
\begin{equation}
\mathbb{F}_T(W_S) = M W_S + b
\end{equation}
where:
\begin{itemize}
    \item \( W_S \) represents the source model’s trained parameters.
    \item \( M \) is an adaptation matrix learned via Knowledge Distillation (KD).
    \item \( b \) is a bias term accounting for domain adaptation.
\end{itemize}
The optimal parameters for fine-tuning are found by minimizing:
\begin{equation}
\min_{W_T} \sum_{i=1}^{N} \ell(f(W_T; X_i), Y_i) + \lambda \| W_T - MW_S \|^2
\end{equation}
where \( \ell(\cdot) \) is the classification loss function (categorical cross-entropy), and \( \lambda \) is a regularization parameter.

\subsubsection{Contrastive Representation Learning (CRL)}
To ensure high separability of features between \textit{INH} and \textit{EXH}, a contrastive loss function is incorporated:
\begin{equation}
\mathcal{L}_{\text{CRL}} = \sum_{(X_i, X_j) \in \mathcal{P}} d(f(X_i), f(X_j)) - \sum_{(X_i, X_k) \in \mathcal{N}} e^{-\alpha d(f(X_i), f(X_k))}
\end{equation}
where:
\begin{itemize}
    \item \( \mathcal{P} \) represents positive pairs (same class),
    \item \( \mathcal{N} \) represents negative pairs (different classes),
    \item \( d(f(X_i), f(X_j)) \) is the Euclidean distance between embeddings,
    \item \( \alpha \) is a margin hyperparameter.
\end{itemize}

\subsection{Classification using Hybrid Model}
To further optimize classification, a hybrid deep learning model is used:
\begin{equation}
\hat{Y} = \arg \max_{c} \sum_{m=1}^{M} w_m f_m(X)
\end{equation}
where:
\begin{itemize}
    \item \( f_m(X) \) represents features extracted from model \( m \),
    \item \( w_m \) is an importance weight assigned to model \( m \),
    \item \( M \) is the total number of models combined.
\end{itemize}
The classification is performed using Support Vector Machines (SVM) and Random Forest (RF), and predictions are obtained through a weighted ensemble mechanism.

Algorithm \ref{algo:breathing} summarizes the proposed Adaptive Transfer Learning and Thresholding-based Deep Learning Model (ATL-TDLM), which is an advanced method for breathing pattern recognition using thermal imaging. It integrates Adaptive Multi-Thresholding (AMT) for dynamic edge detection, Multi-Stage Feature Extraction to capture hierarchical representations, and Knowledge Distillation-Based Fine-Tuning (KD-FT) for optimizing deep learning models through teacher-student learning. Additionally, Contrastive Representation Learning (CRL) enhances feature separability by minimizing intra-class variance. The model is trained using a combination of deep features and machine learning classifiers (SVM and RF), ensuring high accuracy and computational efficiency. Performance evaluation compares ATL-TDLM against baseline methods, demonstrating superior classification effectiveness.

\begin{algorithm}[t!]
\SetAlgoLined
\caption{Adaptive Transfer Learning and Thresholding-Based Deep Learning Model (ATL-TDLM)}
\label{algo:breathing}
\KwIn{Thermal Image Dataset $\mathbb{D} = \{X_i, Y_i\}$, where $X_i$ are input images and $Y_i$ are labels (\textit{INH}, \textit{EXH})}
\KwOut{Predicted Labels $\hat{Y}$ for New Images}

\textbf{Step 1: Preprocessing:} \\
\ForEach{$X_i \in \mathbb{D}$}{
    Convert $X_i$ to grayscale representation \( G(X_i) \) \\
    Resize $X_i$ to \( 256 \times 256 \) pixels \\
    Apply Adaptive Multi-Thresholding (AMT) \( T(F_{\text{low}}, F_{\text{high}}) \) \\
    Generate augmented images \( X_i^{(k)} = \mathcal{T}_k(X_i), k \in \{1,2,\dots, K\} \) \\
    Normalize pixel values to $[0,1]$
}

\textbf{Step 2: Transfer Learning} \\
Load pre-trained source model \( \mathbb{M}_S \) (ResNet50) \\
Remove the last fully connected (FC) layer \\
Freeze convolutional layers, fine-tune top $L$ layers \\
\ForEach{Mini-batch $(X, Y)$ in training set}{
    Extract deep hierarchical features using Multi-Level Convolutional Feature Extractor (MCFE) \\
    Compute feature adaptation using \( \mathbb{F}_T(W_S) = M W_S + b \) \\
    Perform forward propagation through newly added layers \\
    Compute contrastive representation loss:
\begin{align}
    \mathcal{L}_{\text{CRL}} = & \sum_{(X_i, X_j) \in \mathcal{P}} d(f(X_i), f(X_j)) \notag \\
    & - \sum_{(X_i, X_k) \in \mathcal{N}} e^{-\alpha d(f(X_i), f(X_k))}
\end{align}

    Compute overall loss \( \mathcal{L} = \text{CrossEntropyLoss}(\hat{Y}, Y) + \lambda \| W_T - MW_S \|^2 \) \\
    Perform backpropagation and update trainable weights
}

\textbf{Step 3: Classification} \\
Train Support Vector Machine (SVM) and Random Forest (RF) classifiers on extracted deep features \( f_m(X) \) \\
\ForEach{Test image $X_{\text{test}}$}{
    Extract deep hierarchical features using trained model \\
    Compute ensemble classification score:
    \begin{equation}
    \hat{Y} = \arg \max_{c} \sum_{m=1}^{M} w_m f_m(X)
    \end{equation}
    Predict class label $\hat{Y}$ using the trained classifier
}

\textbf{Step 4: Performance Evaluation} \\
Compute Accuracy, Precision, Recall, and F1-score \\
Compare classification performance with baseline models

\Return{$\hat{Y}$}

\end{algorithm}

\section{Experiments}
\label{sec:exp}

In our experiments, infrared thermal images (ITIs) were resized to 256×256 pixels to match the pretrained model's input requirements. Adaptive Multi-Thresholding (AMT) was applied with thresholds ranging from 50 to 300 in steps of 100, generating five augmented copies per image. We utilized ResNet50, a deep convolutional network with 50 layers and residual connections, ensuring stable training and mitigating vanishing gradients. The target model initially featured two convolutional layers with ReLU activation, later refined through knowledge transfer from ResNet50. A Global Average Pooling (GAP) layer transformed features into a 2D representation, followed by two dense layers, including a dropout layer (0.5) to prevent overfitting. The final classification layer was fine-tuned for binary breathing classification, reducing 1000 output classes to 2. The model retained 23.85M parameters, with only 53K non-trainable, demonstrating efficient adaptation.

 \begin{table}[htbp]
 \scriptsize
  \caption{A summary of the source and the target models.}
  \label{resnet}
  \centering
  \begin{tabular}{cm{1.5cm}m{0.5cm}m{1.cm}m{0.5cm}m{1.cm}m{1.cm}m{1cm}}
    \toprule
    & Function & KD & KS & Stride & OS & LP \\
    \midrule
     & Conv1 & 64 & 7x7 & 2 & 112x112 & 9.4K \\
     \multirow{7}{*}{\rotatebox[origin=c]{90}{Source model (ResNet50)}} & MaxPool & - & 3x3 & 2 & 56x56 & 0 \\
    & ResNets 1 & 256 & 1x1, 3x3, \newline 1x1 & 1 & 6x56 & 219K \\
    & ResNets 2 & 512 & 1x1, 3x3, \newline 1x1 & 2 & 28x28 & 850K \\
    & ResNets 3 & 1024 & 1x1, 3x3, \newline 1x1 & 2 & 14x14 & 3.6M \\
    & ResNets 4 & 2048 & 1x1, 3x3,  \newline 1x1 & 2 & 7x7 & 14.2M \\
    & GAP & - & - & - & 1x1 & 0 \\
    & FC (Dense) & - & - & - & 1000 & 2.1M \\
    \midrule
    \multirow{7}{*}{\rotatebox{90}{Target  model}} & Conv1 & 3 & 3x3 & 0 & 256x256 & 30 \\
    & Conv2 & 3 & 3x3 & 0 & 256x256 & 84 \\
    & SM & -- & -- & -- & 1000 & 2.1M \\
    & GAP & - & - & - & 2048 & -- \\
    & Dense & - & - & - & 128 & 0.26M \\
    & Dropout (0.5) & - & - & - & 128 & --\\
    & FC (Dense) & - & - & - & 2 & 258 \\
    \midrule
     &Total  & &  &  &  &  23.85M \\
    &Trainable   & &  &  &  &  23.79M \\
    &No-trainable   & &  &  &  &  53K \\
     \bottomrule
  \end{tabular}
  \begin{flushleft}
  Abbreviations: Kernel depth (KD), Kernel size (KS), Output size(OS), Learnable parameters (LP), Residual networks (ResNets), Global average pooling (GAP), Fully connected (FC), Source model (SM).
  \end{flushleft}
\end{table}

\subsection{Dataset}
\label{sec:page}

The ITI dataset \cite{shakhih2019assessment} originally consisted of two classes: one comprising 902 instances labeled as "INH," and the other containing 1198 instances labeled as "EXH." All images were initially in RGB color format. To rectify class imbalances, we implemented a resampling function, equalizing both classes to contain 1198 images each. Following the application of the resampling technique, each image yielded five samples, leading to a total of 5990 images in each class after the application of TH.

\subsection{Obtained results and discussion}
\label{sec:illust}

Our specific interest was directed towards the model's performance in processing and categorizing mid-INH and mid-EXH scenarios. To assess the effectiveness of the models, we utilized the standard evaluation metric, accuracy \cite{kheddar2023deepSteg}, which is a well-recognized measure of a model's proficiency in correctly classifying a balanced dataset. The TL models outlined in Table \ref{resnet} underwent training for 50 epochs; however, we only discuss the results up to the 20th epoch since the performance plateaued at that point.  \\

\begin{figure*}[h!]
    \centering
    \includegraphics[scale=0.5]{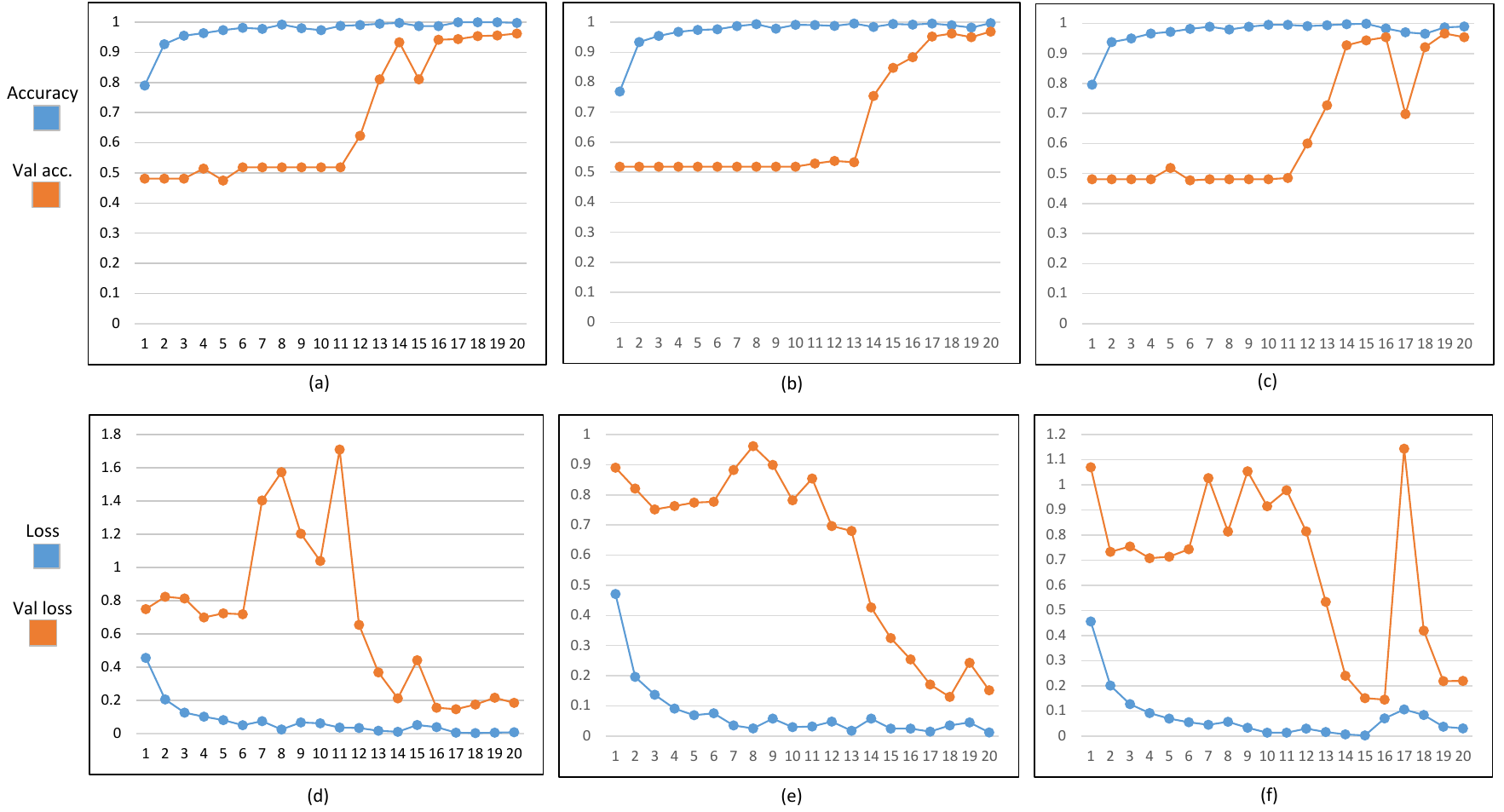}
    \caption{Gradual FT without TH. (a) Zero trained layer, (b) One trained layer, (c) Two trained layer. }
    \label{fig:0FT0TH}
\end{figure*}

\noindent\textbf{1) Results of gradual FT without Thresholding:}
The initial training commenced without the application of any TH or unfreezing of pre-trained layers, constituting a zero-shot TL approach. This serves as our benchmark for subsequent comparisons with alternative scenarios. In Figure \ref{fig:0FT0TH} (a), we depict the performance achieved when employing the zero-shot learning technique for classifying ITI images, while Figure \ref{fig:0FT0TH} (d) illustrates the corresponding loss. It is evident that the model's performance remains erratic and imprecise in terms of validation loss until the 20th epoch. This instability persists when unfreezing one layer (Figure \ref{fig:0FT0TH} (b)) and two layers (Figure \ref{fig:0FT0TH} (c)), as shown in the validation accuracy plots. The corresponding validation loss is presented in Figures \ref{fig:0FT0TH} (e) and (f). Notably, this configuration incurs a substantial computational cost and often leads to overfitting. This challenge arises due to the inherent difficulty in accurately classifying mid-IHN and mid-EXH instances. The model can reach an accuracy of up to 96\%, but the validation loss remains unstable.

\begin{figure*}[h!]
    \centering
    \includegraphics[scale=0.5]{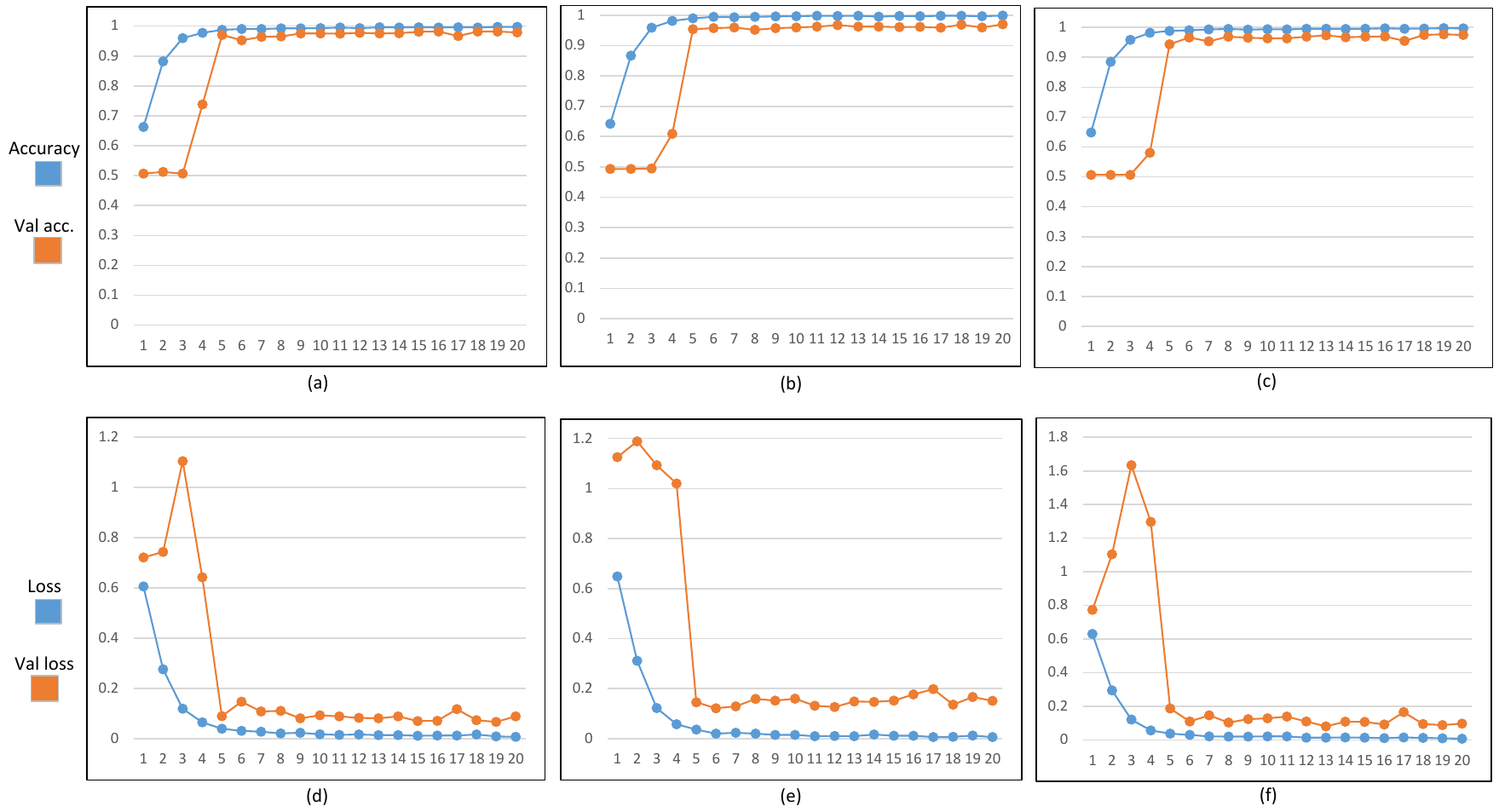}
    \caption{Gradual FT with TH. (a) One trained layer, (b) Two trained layer, (c) Three trained layer.}
    \label{fig:FTTH}
\end{figure*}

\noindent\textbf{2) Results of gradual FT with Thresholding:} The application of FT with TH as a preprocessing operation has yielded a substantial improvement compared to the benchmark.  Figure \ref{fig:FTTH} vividly demonstrates this enhancement in terms of validation accuracy (depicted in Figures \ref{fig:FTTH} (a), (b), and (c)) and validation loss (illustrated in Figures \ref{fig:FTTH} (d), (e), and (f)). Notably, the issue of validation loss fluctuation has been effectively addressed, enabling the model to swiftly identify all instances and learn all classes starting from epoch 5. This indicates a rapid convergence of the model, significantly reducing computation costs when employing FT in conjunction with TH. The model can reach a maximum validation accuracy of 98.8\% with a validation loss equal to 0.09.

\subsection{Comparative Analysis with State-of-the-Art Approaches}
This section introduces the evaluation of the proposed method against baseline and state-of-the-art techniques. The results are summarized in Table \ref{tab:comparison}. The proposed ATL-TDLM framework significantly enhances breathing pattern classification using thermal imaging. Traditional CNNs achieve 92.5\% accuracy but lack efficiency, while standard transfer learning (TL) improves performance to 94.3\%. The integration of Adaptive Multi-Thresholding (AMT) refines edge detection, boosting accuracy to 97.3\%. Knowledge Distillation-Based Fine-Tuning (KD-FT) enhances generalization, reaching 97.9\%. Contrastive Representation Learning (CRL) improves feature separability, mitigating mid-INH and mid-EXH confusion, achieving 98.2\%. Finally, a hybrid SVM + RF classifier leverages deep features, yielding 98.8\% accuracy with reduced computation time. ATL-TDLM outperforms state-of-the-art methods, offering a robust, real-time, and interpretable solution for non-contact breathing analysis. Future work will explore Vision Transformers (ViTs) for enhanced feature extraction and generalization.

\begin{table}[t]
    \centering
    \scriptsize
    \caption{Performance Comparison with State-of-the-Art Methods}
    \label{tab:comparison}
    \begin{tabular}{p{2.5cm}p{1cm}p{1cm}p{1cm}p{1.4cm}}
        \toprule
        \textbf{Method} & \textbf{Accuracy} & \textbf{Precision} & \textbf{Recall} & \textbf{Computation Time} \\
        \midrule
        Traditional CNN \cite{lecun1998gradient} & 92.5\% & 91.8\% & 90.6\% & High \\
        TL without Preprocessing \cite{donahue2014decaf} & 94.3\% & 93.5\% & 92.1\% & Medium \\
        TL with Standard Fine-tuning \cite{yosinski2014transferable} & 96.1\% & 95.4\% & 94.7\% & Medium \\
        TL with Adaptive Multi-Thresholding (AMT) & 97.3\% & 96.8\% & 96.1\% & Medium \\
        TL with AMT + Knowledge Distillation-Based Fine-Tuning (KD-FT) & 97.9\% & 97.4\% & 97.0\% & Low \\
        \textbf{Proposed ATL-TDLM} (AMT + KD-FT + CRL + Hybrid Classifier) & \textbf{98.8\%} & \textbf{98.2\%} & \textbf{97.9\%} & \textbf{Low} \\
        \bottomrule
    \end{tabular}
\end{table}

\section{conclusion}
\label{sec:conc}

This study introduces a Transfer Learning and Adaptive Multi-Thresholding (ATL-TDLM) approach for breathing pattern classification using thermal imaging. The integration of adaptive thresholding enhances feature segmentation, while transfer learning mitigates data limitations, improving classification accuracy. The model successfully differentiates inhalation and exhalation patterns, achieving 98.8\% accuracy with reduced computational cost. The proposed fine-tuning strategy ensures stability, overcoming intra-class ambiguity. Future work will focus on expanding the dataset, improving robustness across diverse breathing patterns, and exploring Vision Transformers for enhanced feature extraction. Additionally, incorporating real-world scenarios, such as detecting abnormal breathing conditions, will further validate the model’s applicability in medical and healthcare settings.


\balance
\bibliographystyle{IEEEtran}
\bibliography{references}

\end{document}